\documentstyle[epsfig,12pt]{article}
\begin{document}
\begin{center}
{\noindent{\bf{Chiral Symmetry Breaking in Gribov's Approach to QCD at Low Momentum}}}

\vspace{0.5cm}

Alok Kumar{\footnote{e-mail address: alok@imsc.res.in}} \\
The Institute of Mathematical Sciences \\
C.P.T.Campus, Taramani Post \\
Chennai 600 113 \\
India. \\
\end{center}

\vspace{3.5cm}  

{\noindent{\it{Abstract}}}

We consider Gribov's equation for inverse quark Green function with and without pion correction. With polar parametrization of inverse quark Green function, we relate the dynamical mass function without pion correction, $M_{0}(q^2)$ and with pion correction, $M(q^2)$ at low momentum. A graph is plotted for $M(q^2)$ and $M_{0}(q^2)$ with q for low momentum. It is found that at  low momenta pion corrections are small.

\vspace{0.5cm}

\newpage 

Gribov [1-5] developed an approach to describe the confinement mechanism and chiral symmetry breaking in QCD, based on the phenomenon of supercritical charges in QCD due to the existence of very light quarks. A mechanism of confinement was given by Gribov  and further elaborated by Ewerz [6]. The related phenomenon chiral symmetry breaking has also been dealt with by Gribov. In this letter, we confine ourselves to the chiral symmetry breaking in Gribov's approach at low momentum transfer.

\vspace{0.5cm}

In most of the studies of chiral symmetry breaking, the Schwinger-Dyson integral equation is used, with suitable approximation methods. Briefly, the one-loop self energy diagram for the quarks gives
\begin{eqnarray}
\Sigma  &=& -i\, C_{F}\, \frac{\alpha_{s}}{\pi}\, \int\, \frac{d^{4}k}{4\pi^2}\,\gamma^{\mu}\,G (k)\,\gamma_{\mu}\,\frac{1}{(q-k)^2},
\end{eqnarray}
in the Feynman gauge for the gluon propagator with $G(k)$ with Green function for the quark, $\alpha_{s}$ as the strong coupling which is taken to be a constant at low momentum region and $C_{F}$ is $\frac{N_{c}^2-1}{2N_{c}}$ where $N_{c}$ is the number of color degree of freedom and for $N_{c}=3$, $C_{F}=\frac{4}{3}$. By differentiating (1) twice with respect to $q_{\mu}$ (external momentum) , one obtains
\begin{eqnarray}
\partial^{2}G^{-1}(q) &=&  C_{F}\,\frac{\alpha_{s}}{\pi}\,\gamma^{\mu}\,G(q)\,\gamma_{\mu}.
\end{eqnarray}
Inclusion of higher order diagram for the most singular contribution essentially is done  when both the derivatives are applied to the same gluon line, and this replces the bare quark-gluon vertices $\gamma_{\mu}$ in (2) by the full vertices  $\Gamma_{\mu}$ which  due to the delta functions appearing in $\partial^{2}$ on the gluon propagator in Feynman gauge are at vanishing gluon momentum i.e. $\Gamma_{\mu}(q,q,0)$. It is to be noted that while the original Schwinger-Dyson equation involves one bare and one full vertex, in Gribov's approach we have two full vertices [7]. The use of the Ward identity
\begin{eqnarray}
\Gamma_{\mu}(q,q,0)  &=& \partial_{\mu}G^{-1}(q),
\end{eqnarray}
then gives
\begin{eqnarray}
\partial^{2}G^{-1}(q)  &=& g\,(\partial^{\mu}G^{-1})\,G\,(\partial_{\mu}G^{-1})\, + ....... \, ,
\end{eqnarray}
where $g = C_{F}\,\frac{\alpha_{s}}{\pi}$ and the dots in (4) stand for less infra-red singular term which are neglected here. In this way, the integral Schwinger-Dyson equation is converted into a partial differential equation for $G(q)$ and this is made possible by the choice of the Feynman gauge. The remarkable feature is that (4) involves only the quark Green function.

\vspace{0.5cm}

The general form of the inverse quark Green function is
\begin{eqnarray}
G^{-1}(q) &=& a(q^2)\,\not\!{q} - b(q^2),  
\end{eqnarray}
where a and  b are two unknown scalar functions of $q^{2}$. A polar parametrization of the Green function in (5) is given by 
\begin{eqnarray}
G^{-1}(q) &=& - \rho\,\exp\left( -\frac{1}{2}\,\phi\,\frac{\not\!{q}}{q}\right),
\end{eqnarray}
where $\rho$ and $\phi$  are functions of $q^2 \,(q = \sqrt{q^{\mu}q_{\mu}})$.
From (5) and (6), we have
\begin{eqnarray}
a(q^2) &=& \frac{1}{q}\,{\rho}\,\sinh\left(\frac{\phi}{2}\right),
\end{eqnarray}
\begin{eqnarray}
b(q^2) &=& {\rho}\,\cosh\left( \frac{\phi}{2}\right).
\end{eqnarray}
The dynamical mass function $M_{0}(q^2)$ of the quark is given by
\begin{eqnarray}
M_{0}(q^2) &=& \frac{b(q^2)}{a(q^2)} = q \, \coth\left( \frac{\phi}{2}\right).
\end{eqnarray}
which involves only $\phi$. The subscript '0' on M will be explained later. Introducing $\xi$ as
\begin{eqnarray}
\xi &\equiv & \ln q = \ln \sqrt{q^\mu q_\mu},
\end{eqnarray}
and denoting $\partial_\xi f(q) = \dot{f}(q)$, the Gribov's equation (4) gets  converted into a pair of coupled differential equations for $\phi$ and $\rho$ as
\begin{eqnarray}
\dot{p} &=& 1 - p^2 - \beta^2 \,\left( \frac{1}{4}\,\dot{\phi^2} + 3 \,\sinh^2\left( \frac{\phi}{2}\right) \right),
\end{eqnarray}
\begin{eqnarray}
\ddot{\phi} + 2 \,p \,\dot{\phi} - 3 \,\sinh\, (\phi) &=& 0  ,
\end{eqnarray}
where 
\begin{eqnarray*}
p &=& 1 + \beta\,\frac{\dot{\rho}}{\rho},
\end{eqnarray*}
with
\begin{eqnarray}
\beta &=& 1 - g = 1- C_F\,\frac{\alpha_s}{\pi} .
\end{eqnarray}
By solving (11) and (12) for $\phi$ and $\rho$ for large  and small q, it was found [2,4] that the dynamical mass function (9) $M_{0}(q^2)$ behaved such that  $M_{0}(0)\neq \,0$. This is a signature for chiral symmetry breaking. In the spontaneous breaking of chiral symmetry, massless pions appear as Goldstone mode in the physical spectrum and they produce correction to the quark propagator. Taking into account this back-reaction of pions on quark, Gribov  obtained 'pion corrected' equation for $G^{-1}(q)$. The coupling of the pion to the quark can be related to the pion decay constant $f_{\pi}$ via Goldberger-Trieman relation, by taking into account the proper isospin factor for light quark flavours. The pion corrected equation for quark Green function, see for review [8] and Gribov [9],
\begin{eqnarray}
\partial^{2} G^{-1} &=& g\,(\partial^{\mu} G^{-1})\,G\,(\partial_\mu G^{-1}) 
- \frac{3}{16 \pi^2 f_\pi^2}\,\{i \gamma_5,G^{-1} \}\,G \,\{i \gamma_5,G^{-1} \} 
\end{eqnarray}
where$f_{\pi}$ is the pion decay constant, 0.093 GeV.
It is to be noted that pion corrected Gribov's equation is still an differential equation involving only light quark's Green function.

\vspace{0.5cm}

Using the same parametrization for this improved equation (14) as in (6) with $(\rho,\phi)$ replaced by $(\rho^{\prime},\phi^{\prime})$, i.e.,
\begin{eqnarray}
G^{-1} &=& - \rho^{\prime} \exp \left( -\frac{1}{2} \phi^{\prime} \frac{\not\!{q}}{q}\right),
\end{eqnarray}
(14) yields,
\begin{eqnarray}
\dot{p^{\prime}} &=& 1 - {p^{\prime}}^2 - \beta^2\, \left( \frac{1}{4} \dot{{\phi^{\prime}}^2} + 3 \sinh^2\left(\frac{\phi^{\prime}}{2}\right)\right) 
\nonumber\\
&&{}
+ \frac{3 \beta q^2}{4\pi^2 f_{\pi}^{2}}\cosh^{2}\left( \frac{\phi^{\prime}}{2} \right), 
\end{eqnarray}
\begin{eqnarray}
\ddot{\phi^{\prime}} + 2\,p^{\prime}\,\dot{\phi^{\prime}} - 3\,\sinh\,(\phi^{\prime}) &=& 0.  
\end{eqnarray}
It is to be observed that the form of the equation for $\phi^{\prime}$ is the same as that for $\phi$ in (12). The dynamical mass with pion correction is
\begin{eqnarray}
M(q^2) &=& {q}\,\coth \left( \frac{\phi^{\prime}}{2} \right).
\end{eqnarray}
For low momentum, $|\overrightarrow{q}|\rightarrow 0$, we linearize the pair of equations (16) and (17) around $(\rho,\phi)$ 
\begin{eqnarray*}
\phi^{\prime} = \phi +  \delta{\phi} \ &\&& \  p^{\prime} = p + \delta{p},
\end{eqnarray*}
and keep only terms  linear in $\delta{\phi}$ and $\delta{p}$. The $\phi$-equations (12) and (17) give the relation
\begin{eqnarray}
2\,\delta{p}\,\dot{\phi} - 3\,\cosh\,(\phi)\,\delta{\phi} &=& 0,
\end{eqnarray}
and  the p-equations (11) and (16) give 
\begin{eqnarray}
\delta{p} &=& \left( -\frac{3\beta^{2}}{4p} + \frac{3\beta q^2}{16 \pi^2 {f_{\pi}}^{2}p} \right)\,\sinh\,(\phi)\,\delta{\phi} + \frac{3\beta q^2}{8 \pi^2 {f_{\pi}}^{2}p}\,\cosh^2 \left( \frac{\phi}{2} \right).
\end{eqnarray}
From (19) and (20), we find
\begin{eqnarray}
\delta{\phi} \,\left[ \frac{\coth\,(\phi)}{2\dot{\phi}} + \frac{\beta^2}{4p} - \frac{\beta \,q^2}{16 \pi^2 {f_{\pi}}^{2}p} \right] &=& \frac{\beta\,q\, M_{0}(q^2)}{16 \pi^2 {f_{\pi}}^{2}p},
\end{eqnarray}
where we have used, $ \coth\left( \frac{\phi}{2}\right) = \frac{M_{0}(q^2)}{q}$ from (9). The dynamical mass with pion correction (18) is
\begin{eqnarray}
M(q^2)&=&q\coth\left(\frac{\phi+\delta\phi}{2}\right),\nonumber\\
&\approx&q \,\left[ \frac{\coth \left( \frac{\phi}{2}\right) + \frac{\delta{\phi}}{2}}{1 + \frac{\delta\phi}{2}\coth\left(\frac{\phi}{2}\right)}\right],\nonumber\\
&\approx&q\,\left[\coth\left(\frac{\phi}{2}\right)+\frac{\delta{\phi}}{2}\right]\left[1-\frac{\delta{\phi}}{2}\coth\left(\frac{\phi}{2}\right)\right],\nonumber\\
&\approx&q\left[\coth\left(\frac{\phi}{2}\right)+\frac{\delta{\phi}}{2}\left(1-\coth^2\left(\frac{\phi}{2}\right)\right) \right],\nonumber
\end{eqnarray}
where we have kept the terms linear in $\delta\phi$. We use the  relation $ \coth\left( \frac{\phi}{2}\right) = \frac{M_{0}(q^2)}{q}$ from (9), and $M(q^2)$ becomes,
\begin{eqnarray}
M(q^2) &=& M_{0}(q^2)+q\left(\frac{\delta\phi}{2}\right)\left(1-\frac{M_{0}^2(q^2)}{q^2}\right),
\end{eqnarray}
substituting  $\delta{\phi}$ from (21) in (22), we find
\begin{eqnarray}
M(q^2) &=& M_{0}(q^2)\left[ 1+\left(\frac{\beta q^2}{32\pi^2 f_{\pi}^2p}\right)\left(\frac{1}{\alpha}\right)\left(1-\frac{M_{0}^2(q^2)}{q^2}\right)\right],
\end{eqnarray}
where
\begin{eqnarray*}
\alpha &=& \left[ \frac{\coth\,(\phi)}{2\dot{\phi}} + \frac{\beta^2}{4p}-\frac{\beta q^2}{16\pi^2 f_{\pi}^2 p}\right].
\end{eqnarray*}
Equation (23) gives a relationship between dynamical mass of quarks  with pion correction, $M(q^2)$ and without pion correction, $M_{0}(q^2)$ at low momentum. This is our main result. Further the expression in (23) and $\alpha$ involve solutions to (11) and (12). It can be seen from equation (23) that in the limit $f_{\pi}\rightarrow \infty$ (i.e. no pion correction ) $M(q^2) \rightarrow M_{0}(q^2)$.

\vspace{0.5cm}

Now we consider the solutions of (11) and (12) in the infrared region $q \rightarrow 0$. In [6], one possible solution when $|\overrightarrow{q}|\rightarrow 0$ is $p \rightarrow p_{0}$ with $p_{0}^2=1$ and $\phi = C \, e^{\xi}$ for $p_{0}=1$, the arbitrary constant C has the dimension inverse of length.
We use the expansion for $\coth(x)$ and keep first three terms only [10],
\begin{eqnarray}
\coth(x) &\approx& \frac{1}{x} + \frac{x}{3} - \frac{x^3}{45}.
\end{eqnarray}
Using the solution for $\phi$ at low momentum $\phi = C q$  and $\dot{\phi} = C q$, the dynamical mass without pion correction $M_{0}(q^2)$ is given by,
\begin{eqnarray}
M_{0}(q^2)  &=&   \frac{2}{C} + \frac{C\,q^2}{6} - \frac{C^3\,q^4}{360},
\end{eqnarray}
and $\alpha$ is given by
\begin{eqnarray}
\alpha &=& \frac{1}{2}\,\left(\frac{1}{C^2\,q^2}+\frac{1}{3}-\frac{C^2\,q^2}{45}\right)\,+\,\frac{\beta^2}{4}-\frac{\beta\,q^2}{16\,\pi^2\,f_{\pi}^2},
\end{eqnarray}
and the dynamical mass with pion correction $M(q^2)$ is given by
\begin{eqnarray}
M(q^2)&=&M_{0}(q^2)\left[1+\frac{\beta C^2 q^2\frac{\left(q^2-M_{0}^2(q^2)\right)}{16\pi^2 f_{\pi}^2}}{1+(\frac{1}{3}+\frac{\beta^2}{2})C^2\,q^2-\frac{C^4\,q^4}{45}-\frac{\beta\,C^2\,q^4}{8\,\pi^2 \,f_{\pi}^2}}\right].
\end{eqnarray}
This is valid in the low momentum region only. In the limit $q\rightarrow 0$, we find $M(0)\rightarrow M_{0}(0)=\frac{2}{C}\neq 0$. For space-like momenta we replace '$q^2$' by '-$q^2$', and equations (25) and (29) change to,
\begin{eqnarray}
M_{0}(q^2)  &=&   \frac{2}{C} - \frac{C\,q^2}{6} - \frac{C^3\,q^4}{360},
\end{eqnarray}
\begin{eqnarray}
M(q^2)&=&M_{0}(q^2)\left[1+\frac{\beta\,C^2\,q^2\frac{\left(q^2+M_{0}^2(q^2)\right)}{16\pi^2 f_{\pi}^2}}{1-(\frac{1}{3}+\frac{\beta^2}{2})C^2\,q^2-\frac{C^4\,q^4}{45}-\frac{\beta\,C^2\,q^4}{8\,\pi^2 \,f_{\pi}^2}}\right].
\end{eqnarray}
We use (28) and (29) to exhibit the behaviour of $M_{0}(q^2)$ and $M(q^2)$ at low momentum. We use $f_{\pi}=0.093\,GeV$ [11] and the arbitrary constant C is taken to reproduce the numerical value of $M_{0}(0)=M(0)$ as estimated in [6]. In [6], $M_{0}(0)=M(0)=0.1\,GeV$ and so from (25), $C = 20GeV^{-1}$. At low momenta we take the strong coupling constant to be constant and use the supercritical value $\alpha_{c}=0.43$ as found in [2,4] by Gribov. For this value of  $\alpha_{s}$, $\beta =1-g\,= 0.8175\,$. We plotted the variation of $M_{0}(q^2)$ and $M(q^2)$ for q\,=\,0 to 0.045 GeV using Matlab. The output graph is given in Figure1 and the solid line  corresponds to variation of $M_{0}(q^2)$ and the broken line is for $M(q^2)$. It is found from the Figure1 that in the low momentum region pion correction to quark's mass is small. This feature is  similar to the study of [6] at  large momentum.
 
\vspace{0.5cm}

{\noindent{\bf{Acknowledgements}}}

\vspace{0.5cm}

I thank Prof. R. Parthasarathy (IMSc, Chennai and CMI, Chennai) for showing the problem and providing constant help and encourgement during the completion of this work. The award of JRF fellowship by IMSc is acknowledged with thanks.

\vspace{0.5cm}

\newpage

{\noindent{\bf{References}}}

\vspace{0.5cm}

\begin{enumerate}
\item V.\,N.\,Gribov, Phys. Scripta {\bf{T 15}} (1987) 164.
 \item V.\,N.\, Gribov, {\sl Possible Solution of the Problem of Quark Confinement}, Lund preprint LU-TP 91-7 (1991).
\item V.\,N.\ Gribov, {\it Eur.\ Phys.\ J. }{\bf C 10} (1999) 91 [hep-ph/9902279].
\item V.\,N.\ Gribov, {\it Eur.\ Phys.\ J. }{\bf C 10} (1999) 71 [hep-ph/9807224].
\item V.\,N.\ Gribov, Orsay lectures on confinement (I-III):\\
LPT Orsay 92-60, hep-ph/9403218;\\
LPT Orsay 94-20, hep-ph/9404332;\\
LPT Orsay 99-37, hep-ph/9905285.
\item Carlo Ewerz, {\sl Gribov's Equation for the Green Function of Light Quarks}, Eur. Phys. J. {\bf{ C 13}} (2000) 503-518 [hep-ph/0001038].
\item Carlo Ewerz, {\sl Gribov's Picture of Confinement and Chiral Symmetry Breaking}, Talk presented at Gribov-75 Memorial Workshop, Budapest, May 2005 [hep-th/0601271].
\item Yu.\,L. Dokshitzer, and D.E. Kharzeev, {\sl The Gribov Conception of Quantum Chromodynamics}, Ann. Rev. Nucl. Part. Sci. {\bf{ 54}} (2004) 487-524 [hep-ph/0404216].
\item V.\,N.\ Gribov, {\sl The Gribov Theory of Quark Confinement }, Ed. J. Nyiri, World Scientific Publication, 2001.
\item Alan Jeffrey,\,Handbook of Mathematical Formulas and Integrals,\,Elsevier Academic Press,\,Third Edition\,(2004).
\item M.\,E.\,Peskin and D.\,V.\,Schroeder,\," An Introduction to Quantum Field Theory",\,Westview Press,\,(1995).
\end{enumerate}
\begin{figure}
\begin{center}{\hbox{\epsfig{figure=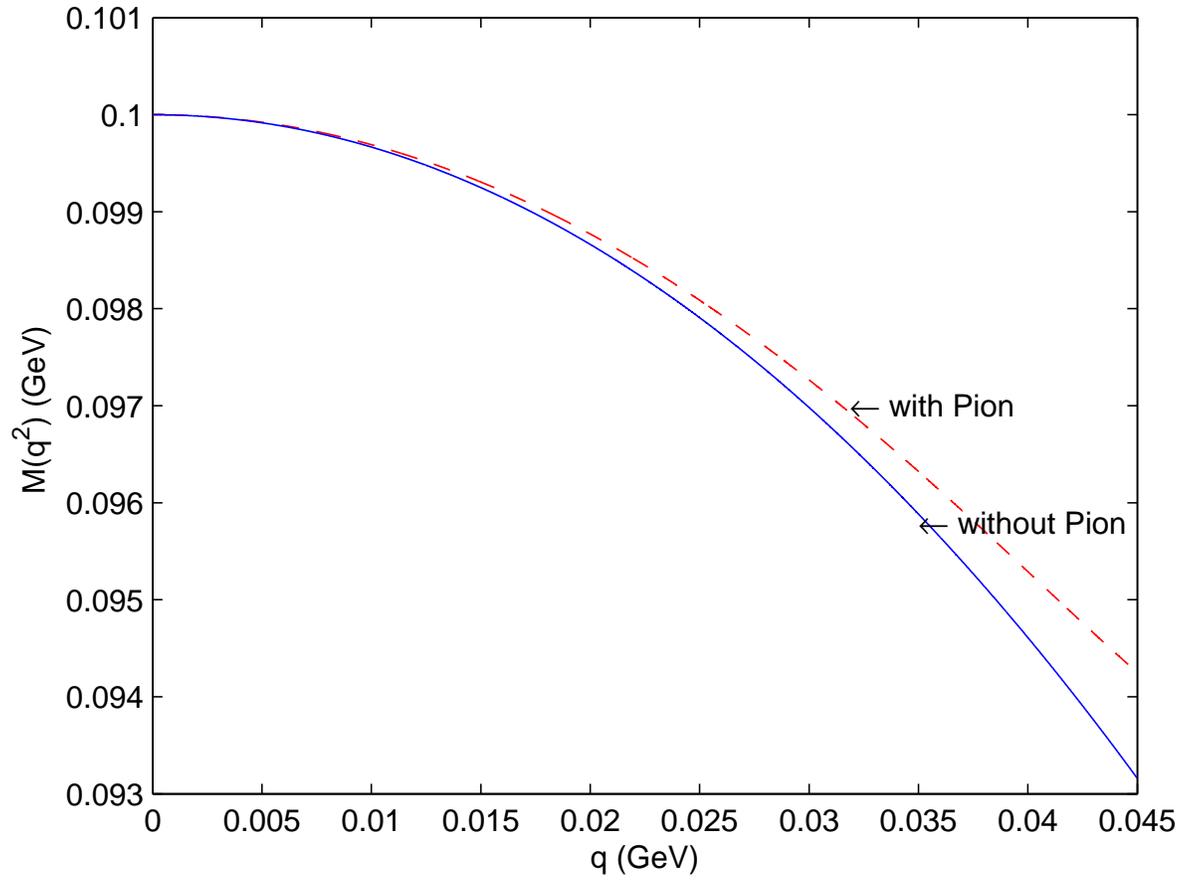,height=5in}}}
\end{center}
\caption{Variation of Dynamical Mass  Function with Pion Correction, $M(q^2)$ (broken line)  and without Pion Correction, $M_{0}(q^2)$ (solid line) with Momentum Scale }
\end{figure}

\end{document}